\documentclass[12pt]{amsart}

\usepackage{amssymb,amsmath,amscd,amsthm}

\usepackage{euscript}
\usepackage{multicol}
\usepackage{amsmath}
\usepackage{times}
\usepackage[active]{srcltx}

\newtheorem{theorem}{Theorem}[section]
\newtheorem{proposition}{Proposition}[section]
\newtheorem{corollary}{Corollary}[section]
\newtheorem{lemma}{Lemma}[section]
\numberwithin{equation}{section}

\date{}

\begin{document}

\date{}

\title[Absolutely continuous and pure point  spectra]{Absolutely continuous and pure point  spectra of discrete  operators with sparse  potentials}
\author{S.  Molchanov, O. Safronov, and B. Vainberg }
\address{ S.  Molchanov: Dept. of Math. and Statistics, UNCC and Higher School of Economics, Russia; O. Safronov, and B. Vainberg: Dept. of Math. and Statistics, UNCC. }
\email{smolchan@uncc.edu, osafrono@uncc.edu, brvainbe@uncc.edu }
\maketitle

\begin{abstract}We consider the discrete Schr\"odinger operator $H=-\Delta+V$ with a sparse potential $V$ and find conditions  guaranteeing either existence of  wave operators  for the pair $H$
and $H_0=-\Delta$,
or presence of  dense purely point  spectrum  of the operator $H$ on some interval $[\lambda_0,0]$  with $\lambda_0<0$.
\end{abstract}

\section{introduction}
The class of  sparse potentials  was introduced in the spectral theory of  one-dimensional Schr\"odinger operators by D. Pearson \cite{Pearson}.
Such   potentials  are intermediate  between  compactly supported (or  fast  decaying) potentials associated  to the scattering  theory,   and the   generic  bounded potentials
( such as periodic, almost periodic, random ergodic  potentials)  which are  studied  in the solid state physics.

Typical   sparse potentials have the form
\begin{equation}\label{sparseVdef}
V(x)=\sum_{n\in{\Bbb N}} a_n \phi(x-x_n)
\end{equation}
where $\phi\in C_0^\infty$ is  a  fixed  bump function and $x_n$  is  a  sequence  for which  the quantity
$d(n)={\rm dist}\, (x_n, \bigcup_{j\neq n}\{x_j\})$    grows  as $n\to\infty$. The amplitudes $a_n$ either  slowly  tend to zero, or are of order $O(1)$ (as in the case   where  they are independent  identically distributed  random variables).

The paper  by Kiselev, Last and Simon \cite{KLS} as well as   the paper  by  Molchanov \cite{Molch} extend  the results of \cite{Pearson} in  different  directions:

Let $H=-\frac{d^2}{dx^2}+V$  be the operator  with a  potential of the form \eqref{sparseVdef} acting   in the space $L^2[0,\infty)$.
Assume  that the  boundary condition at the origin   is   the Dirichlet  condition. Suppose also that  $\phi\geq 0$,  the sequence $x_n$ is monotone and $a_n\to 0$ as $n\to\infty.$

1) If $\sum_n a_n^2<\infty$  and $x_n/x_{n+1}=o(1)$ as $n\to\infty$, then  the spectrum is pure absolutely  continuous on $(0,\infty)$.

2)  If $\sum_n a_n^2=\infty$  and $x_n/x_{n+1}=o(1)$ as $n\to\infty$, then  the spectrum is purely singular continuous on $(0,\infty)$.

These statements   could be compared  with the results    of  the paper by Kotani and Ushiroya \cite{KU}
where  the  authors    study  the transition from the absolutely  continuous  spectrum to  the  pure point  spectrum
for the  operator with a random potential.

Technically, the analysis  of   the absolutely continuous spectrum   of the operator $H$   under  the condition $\sum_n a_n^2<\infty$
resembles  the  theory of lacunary  Fourier  series and  is  based on the  idea of the stochastization  of the  phase of the solution  of the equation $H\psi=k^2\psi$, $k\in {\Bbb R}\setminus \{0\}$,
as  one passes  from one  bump to  another.

One should also  notice  that  the spectral properties of   the  Hamiltonian $H$  depend on the nature of the  elementary bump $\phi$.
For  instance,  if  all $a_n=1$  and $\phi$  is a reflectionless  potential, then the spectrum   of $H$ is  pure  absolutely  continuous on $(0,\infty)$
provided the  sequence $\{x_n\}$  is exponentially sparse (see \cite{Molch}).  At the same time, if the reflection coefficient $r(k)$ constructed  for   the   bump $\phi$ is not identically zero,
then the  spectrum of $H$  is  purely singular continuous  on $(0,\infty)$ provided  some additional mild technical assumptions are fulfilled (see \cite{Molch}).

One expects   the multi-dimensional spectral theory of Schr\"odinger operators with   sparse potentials
to  be  different from the one-dimensional theory, because the  waves "reflected"  from one individual  bump  decay at infinity, if $d\geq 2$.
More precisely, the   solution of the equation $-\Delta \psi+a_n\phi(x-x_n) \psi=k^2\psi$ constructed  for one bump  decays  as $O(|x|^{-(d-1)/2})$  as $|x|\to\infty$.
The latter leads  to  the  fact that the absolutely  continuous spectrum of the milti-dimensional  Schr\"odinger operator with a sparse potential can  cover  the
positive half-line  even in the case  where the coefficients  $a_n$ do not decay.

Indeed,  one of the results of the paper by Safronov   \cite{Safronov}
says  that  if $V_\xi$ is  a function of the form
\[
V_\xi(x)=\sum_{n\in {\Bbb Z}^d} a_n \xi_n \chi(x-n),
\]
where $\xi_n$   are  independent identically distributed  bounded   random variables  such that ${\Bbb E}(\xi_n)=0$  and $\chi$   is the characteristic  function of the cube $[0,1)^d$,
then  the absolutely continuous   spectrum of the  operator
$
-\Delta+t V_\xi
$
almost surely covers  the positive half-line  for almost every $t\in {\Bbb R}$  provided
\begin{equation}\label{l2uslovie}
\sum_{n\neq 0}\frac{a_n^2}{|n|^{d-1}}<\infty.
\end{equation}
This statement should be  compared  with the  results of  Bourgain \cite{Bourgain} and Denisov \cite{Den}  in which  the authors  assume  that $|a_n|\leq C(1+|n|)^{-1/2-\varepsilon}$ with $\varepsilon>0$ and  prove the corresponding  claim for every $t$.
The  main difference  between  \cite{Bourgain}  and \cite{Den}   is that the first paper deals   with  the  discrete operator on  the lattice ${\Bbb Z}^2$   while the second one  handles the  continuous operator on ${\Bbb R}^d$.
The potentials  considered in the paper  \cite{KKO} are  also decaying  at infinity.

One should  mention  that  the corresponding  statement about  the absolutely continuous spectrum of the  discrete  Schr\"odinger operator on ${\Bbb Z}^d$ under  the condition similar to \eqref{l2uslovie}  has not been proved.
Neither has been  proved  a discrete analogue   the Laptev-Naboko-Safronov (see \cite{LNS})  theorem saying that if  $V\geq0$  and
\begin{equation}\label{LNSc}
\int_{{\Bbb R}^d}  \frac { V(x)}{(1+|x|)^{d-1}}dx<\infty,
\end{equation}
then the absolutely  continuous spectrum of the operator $-\Delta+V$  covers the   half-line $[0,\infty)$.
Note that both conditions \eqref{l2uslovie}  and \eqref{LNSc}  are fulfilled    for some  sparse potentials that  do not  decay at infinity.

The operator $H$ with a  sparse potential that does not decay at infinity can  have  pure  point  spectrum outside of the spectrum of the  free Laplace operator.
As a result, one can expect the discrete operator $H$  with a negative  random sparse potential  to have absolutely  continuous spectrum on $[0,4d]$  and dense pure point spectrum on some  interval $[-a,0)$  with $a>0$.
This phenomenon was proved in the paper \cite{Molch2}  by Molchanov    but   without  the specification of the physical nature of the  absolutely continuous  component  of the spectrum.  In the present paper, we  develop   the scattering theory  for  a  class of  discrete operators on the lattice ${\Bbb Z}^d$ with sparse potentials $V$  and
prove existence of  wave operators.
\section{Statement of the main results}

Let $H_0=-\Delta$ be the  "free" operator on the lattice ${\Bbb Z}^d$  whose  action is  defined by the formula
\[
\bigl[H_0u\bigr](n)=\sum_{|n-j|=1}\Bigl(u(n)-u(j)\Bigr).
\]
Let $H=-\Delta+V$, where $V$  is the operator of multiplication  by a  bounded  real-valued  function $V:{\Bbb Z}^d\to {\Bbb R}$. We are interested in  the question of  existence of the wave operators
\begin{equation}\label{1}
\text{s -}\lim_{t\to\pm\infty}e^{-itH}e^{itH_0}=:W_\pm.
\end{equation}
\begin{theorem}\label{main1}
Assume that
\[
\sum_{n\in {\Bbb Z}^d\setminus \{0\}}\frac{|V(n)|}{|n|^{(d-1)/2}}<\infty.
\]
Then the wave operators \eqref{1}  exist.
\end{theorem}
Note  that Theorem~\ref{main1} is applicable to  operators with  sparse potentials, for which the quantity
 \begin{equation}\label{definitiond1}d(n)=\text{dist}\,\bigl(n, \,\, \text{supp}(V)\setminus \{n\}\bigr)\end{equation}
tends to infinity   sufficiently fast as $|n|\to\infty$.
Another  theorem that deals  with sparse potentials  is the followings statement, in which $\chi_n$ is the characteristic  function of the one point set $\{n\}\subset {\Bbb Z}^d$.

\begin{theorem} \label{main2} For  an arbitrary   number  $\lambda_0<0$, let
\[
a=\frac1{\bigl( (H_0-\lambda_0)^{-1}\chi_0,\chi_0\bigr)}.
\]
Let $\Omega$  be a  fixed subset of the lattice ${\Bbb Z}^d$, such that
\[
V_\xi(n)=\begin{cases}
\xi_n,\qquad \text{if} \qquad n\in \Omega,\\
0,\qquad \text{if} \qquad n\notin \Omega,
\end{cases}
\]
where $\{\xi_n\}_{n\in {\Bbb Z}^d}$ are independent  random variables  uniformly distributed  on the interval $[-a,0]$.
 Assume that
that the   quantity $d(n)$ defined  by  \eqref{definitiond1} obeys  the condition
\[
\lim_{ |n|\to\infty, \, n\in \Omega}\frac{d(n)}{|n|^\delta}=\infty
\] for some $\delta>0$.
Then the operator  $H_\xi=-\Delta+V_\xi$ almost surely  has a  dense  pure point spectrum in  the interval $[\lambda_0,0)$.
\end{theorem}

According to Theorems~\ref{main1}  and ~\ref{main2},  there  are discrete Schr\"odinger operators  whose absolutely continuous spectrum  "fills" the interval $[0,4d]$ while their
 pure point spectrum is dense in  $[\lambda_0,0)$.

\section{\ Proof of Theorem~\ref{main1}.}

{\bf Standard notations}. For a  closed  linear operator $T$, the symbols ${\mathcal D}(T)$, $\sigma(T)$ and  $\sigma_p(T)$ denote its   domain, its spectrum and the  set of eigenvalues (which does not have to be closed).  If $T$  is self-adjoint, then $E_T(\cdot)$  denotes  the operator-valued spectral measure of the operator $T$.

\bigskip

Besides the standard  wave operators $W_\pm$, we will  consider the modified operators $W_\pm(J)$  defined  by
\begin{equation}\label{2}
W_\pm(J)=\text{s -}\lim_{t\to\pm\infty}e^{-itH}J e^{itH_0},
\end{equation}
 where $J:
\ell^2({\Bbb Z}^d)\to\ell^2({\Bbb Z}^d)$  is assumed  to be bounded.
We  will employ one   of the statements  of the scattering  theory,  using  the notion of a trace class operator.
One  says that an  operator $T$ on a separable Hilbert  space belongs to the  class ${\frak S}_p$,
if the sequence $\{s_j(T)\}$ of singular values of this operator is an $\ell^p$-sequence.
In this  case, the norm of $T$  in ${\frak S}_p$ is defined as the norm of
the sequence $\{s_j(T)\}$ in $\ell^p$.
The classes ${\frak S}_1$  and ${\frak S}_2$  are called the trace class and the Hilbert-Schmidt  class,  correspondingly.

One of  theorems  by Pearson (see \cite{Yafaev}) leads to the following result:
\begin{theorem}
Let $E_0(\cdot)$  and $E(\cdot)$   be the operator-valued spectral measures  of $H_0$ and $H$  correspondingly.
Assume that  the operator
\begin{equation}\label{3}
E(a,b)\bigl(HJ-JH_0\bigr)E_0(a,b)\in {\frak S}_1
\end{equation}
is a trace class operator  for  any bounded interval $(a,b)$. Then  the  limits $W_\pm(J)$ in \eqref{2} exist.
\end{theorem}
We will apply this theorem in the case where $J$ commutes  with $H_0$.
In this case, the condition \eqref{3} turns into the relation
\begin{equation}\label{4}
E(a,b)VJ E_0(a,b)\in {\frak S}_1,
\end{equation}
while  the limit \eqref{2}  coincides  with the limit
\[
\text{s -}\lim_{t\to\pm\infty}e^{-itH}e^{itH_0}J.
\]
The latter observation implies the following lemma.
\begin{lemma}\label{lemma3.1}
Let  $\{J_l\}_{l\in {\Bbb N}}$ be a family of bounded operators commuting with $H_0$  and such that
\begin{equation}\label{J}
\text{s-}\lim_{l\to\infty}J_l=I.
\end{equation}
Assume that
\[
VJ_l E_0(a,b)\in {\frak S}_1
\]
for  any  bounded  interval $(a,b)$  and  for any $l$. Then  the limits $W_\pm$ in \eqref{1}  exist.
\end{lemma}

\bigskip

{\bf Remark.}
Replacement of  \eqref{J}  by  the  assumption \[
\text{s-}\lim_{l\to\infty}J_l=J.
\]  leads to  the existence of  the limits \eqref{2}.

\bigskip

In order to   construct  operators $J$  needed  for our purposes, one has to introduce the operator
$\Phi: L^2({\Bbb T}^d)\to \ell^2({\Bbb Z}^d)$  setting
\[
\bigl[\Phi u\bigr](n)=\frac1{(2\pi)^{d/2}}\int_{{\Bbb T}^d} e^{-i\xi n} u(\xi)d\xi,\qquad {\Bbb T}^d=[0,2\pi)^d.
\]
It is  very well known that
\begin{equation}\label{diagonalizaciyaH}
H_0=\Phi \bigl[a\bigr] \Phi^{-1},
\end{equation}
where $[a]$  denotes  the operator of multiplication by the  function \[a(\xi)=\sum_{j=1}^d(2-2\cos(\xi_j)),\qquad \xi=(\xi_1,\xi_2,\dots,\xi_d),\] called the symbol of the operator $H_0$.

\bigskip

{\bf Definition of a regular set.} An  open set $\Omega\subset {\Bbb T}^d$  is called  regular, if

(1) $\nabla a(\xi)\neq 0,\qquad \forall \xi\in \Omega$.

(2)  The curvatures  of the  level surfaces $a(\xi)=const $  are different  from zero at  all   points $\xi\in \Omega$.

(3)  $\Omega$ is  diffeomorphic  to a set $[\alpha,\beta]\times U$, where $U$ is an open set in ${\Bbb R}^{d-1}$. Moreover, the corresponding diffeomorphism
\[
\tau:\Omega\to [\alpha,\beta]\times U
\]
is a function $\tau=(\tau_1,\tau_2,\dots,\tau_d)$  having the property $\tau_1(\xi)=a(\xi),\quad \forall \xi\in \Omega$.
 Without  loss  of  generality, we will assume that the diffeomorphism $\tau$ can be extended into a  larger   domain $\tilde \Omega$ containing  $\Omega$, so that the  corresponding Jacobian
is a  bounded separated  from zero function on $\Omega$.

\bigskip

Note that for  any $\varepsilon>0$, there is a  disjoint collection of sets $\{\Omega_j\}_{j=1}^N$
having the properties (1)-(3) such that  the Lebesgue measure  of  the set
\[
{\Bbb T}^d\setminus \bigcup_{j=1}^N \Omega_j
\]  is  smaller than $\varepsilon$. Therefore,
the collection   of orthogonal projections $P_{\Omega_j}$ in $L^2({\Bbb T}^d)$ onto  the  space of  functions   vanishing outside of  $\Omega_j$ is a   family of operators  with the property
\[
\sum_j P_{\Omega_j}f\to f,\qquad \text{as}\quad \varepsilon\to 0,\quad \forall f\in L^2({\Bbb T}^d).
\] Consequently, we obtain  the  following statement:

\begin{corollary}\label{corollaryz3.1}
Let  $J_\Omega: \ell^2({\Bbb Z}^d)\mapsto \ell^2({\Bbb Z}^d)$  be the  operator
\[
J_\Omega=\Phi \chi_\Omega \Phi^{-1}
\]
where $\chi_\Omega$  is the  characteristic  function of  a set $\Omega\subset {\Bbb T}^d$. If the limits $W_\pm(J_\Omega)$  defined  by \eqref{2}  with $J=J_\Omega$
exist  for all  sets $\Omega$  having the properties (1)-(3), then the wave operators $W_\pm$  defined in \eqref{1}  also exist.
\end{corollary}

The next  step in our  arguments  will be  an approximation of  the operators $J_\Omega$
 by  operators $J_{\Omega,n}$  commuting with $H_0$ and having the property that
\[
J_{\Omega,n}\,f\to J_\Omega \,f,\qquad \text{as } \quad n\to\infty,\quad   \forall f \in \ell^2({\Bbb Z}^d).
\]
Obviously, existence of  the limits \eqref{2}   with  $J=J_{\Omega,n}$ would  imply  existence of  operators
$W_\pm(J_\Omega)$.
Therefore,  according to Corollary~\ref{corollaryz3.1}, the wave operators  $W_\pm$  would also exist.

Note that, for any $f\in \ell^2({\Bbb Z}^d)$, the  element   $J_\Omega f\in \ell^2({\Bbb Z}^d)$  is  the sequence
\begin{equation}\label{vainberg1}
\bigl[J_\Omega f\bigr](j)=\frac1{(2\pi)^{d/2}}\int_{[\alpha,\beta]\times U}e^{i\,j\cdot \xi(\tau)} \omega(\tau) \bigl[\Phi^{-1}f\bigr](\xi(\tau))d\tau,\quad j\in {\Bbb Z}^d,
\end{equation}
where $\xi(\cdot)$  is  the  inverse of the mapping $\tau(\cdot)$ and
$\omega(\tau)=\bigl|\frac{\partial \xi}{\partial \tau}\bigr|$  is its Jacobian.

Let $\{\phi_k\}_{k=1}^\infty$ be an orthonormal basis  in $L^2(U)$. Without loss  of  generality, one can assume that
$\phi_k\in C_0^\infty(U)$.  We define  the orthogonal projections   $P_n:L^2([\alpha,\beta]\times U)\to L^2([\alpha,\beta]\times U)$ setting
\[
\bigl[P_n u\bigr](\tau)=\sum_{k=1}^n \phi_k(\tau') \int_U u(\tau_1,\eta)\bar \phi_k(\eta)d\eta,\qquad \tau=(\tau_1,\tau')\in [\alpha,\beta]\times U.
\]
After  that, we define  the operators $J_{\Omega,n}$ by
\begin{equation}\label{vainberg2}
\bigl[J_{\Omega,n} f\bigr](j)=\frac1{(2\pi)^{d/2}}\int_{[\alpha,\beta]\times U}e^{i\,j\cdot \xi(\tau)} \omega(\tau)
\Bigl[P_n \bigl[[ \Phi^{-1}f](\xi(\cdot))\bigr]\Bigr](\tau)\, d\tau,\quad j\in {\Bbb Z}^d.
\end{equation}
Taking into account the fact that $a(\xi(\tau))=\tau_1$  for all $\tau\in[\alpha,\beta]\times U$, that is,  the symbol of $H_0$
coincides with the  first of   the $\tau$-coordinates,
we obtain  from  \eqref{diagonalizaciyaH} and \eqref{vainberg2} that  $\bigl[P_n ([ \Phi^{-1}H_0 f]\circ \xi)\bigr](\tau)=\tau_1\bigl[P_n ([ \Phi^{-1}f]\circ  \xi)\bigr](\tau)$.
Hence, the operators  $J_{\Omega,n}$  commute with $H_0$.
Furthermore, since  the sequence  $P_n$  converges to $I$ strongly in $L^2([\alpha,\beta]\times U)$, it converges to the same limit in the weighted space $L^2([\alpha,\beta]\times U, \omega)$, because the norms in these two spaces are equivalent.
Comparing \eqref{vainberg1}  and \eqref{vainberg2}, we conclude that $J_{\Omega,n}\to J_\Omega$ strongly as $n\to\infty$.

Thus, it remains to establish  existence of  limits \eqref{2} for $J=J_{\Omega,n}$.
The latter follows  from the statement below.

\begin{proposition}
Let $\chi_j$ be the  characteristic function of  the one-point set $\{j\}\subset {\Bbb Z}^d$. Let $J_{\Omega,n}$
 be the   operators  defined  above. Then there  is a positive constant $C=C(d,\Omega,n)$ depending  only
on $d,\Omega$  and  the choice of the collection  $\{\phi_k\}_{k=1}^n$  such that
\begin{equation}\label{5}
\|V\chi_j J_{\Omega, n} \| \leq C\frac{|V(j)|}{1+    |j|^{(d-1)/2}    },\qquad \forall j\in {\Bbb Z}^d.
\end{equation}.
\end{proposition}
{\it Proof.}
The relation  \eqref{5} follows  from the estimate
\begin{equation}\label{ShaV}
Q_k(\tau_1, j):=\Bigl|\int_{U}e^{i j\cdot \xi(\tau_1,\tau')}\phi_k(\tau')\omega(\tau_1,\tau') d\tau' \Bigr|\leq \frac{C_{\tau_1}}{1+    |j|^{(d-1)/2}    },\qquad \forall j\in
 {\Bbb Z}^d,
\end{equation}
which holds  for each  fixed $\tau_1\in [\alpha,\beta]$.  Such inequalities were studied  systematically in the paper \cite{SV} by Shaban and  Vainberg.
In particular, it was  shown that $C_{\tau_1}$ is  a bounded  function of $\tau_1$ on $[\alpha,\beta]$.
To prove \eqref{5}, one needs to observe that, for any $u\in L^2({\Bbb T}^d)$,
\begin{equation}\label{ShaV2}
\Bigl|\bigl[V J_{\Omega, n} \Phi  u\bigr](j)\Bigr|\leq (2\pi)^{-d/2}|V(j)|\sum_{k=1}^n \int_\alpha^\beta \,Q_k(\tau_1,j) g_k(\tau_1)d\tau_1,
\end{equation}
where
\begin{equation}\label{ShaV3}
g_k(\tau_1)=\Bigl|\int_U u\bigl( \xi (\tau_1,\tau')\bigr) \bar \phi_k (\tau') d\tau'\Bigr|\leq
\Bigl(\int_U | u\bigl(\xi(\tau_1,\tau')\bigr)|^2d\tau'\Bigr)^{1/2}.
\end{equation}
The  inequality \eqref{ShaV3} implies that $\int_\alpha^\beta g_k(\tau_1)d\tau_1\leq C\|u\|_{L^2({\Bbb T}^d)}$.
Combining this estimate with the relation \eqref{ShaV}, we  infer \eqref{5}   from \eqref{ShaV2}. $\,\,\,\,\,\Box$

\bigskip

Since $V\chi_j J_{\Omega, n}$  is a rank one operator, its  trace  class  norm coincides  with the  usual norm.
Consequently, we obtain:
\begin{corollary}\label{corollary3.2}
Assume that
\[
\sum_{j\in {\Bbb Z}^d\setminus\{0\}}\frac{|V(j)|}{  |j|^{(d-1)/2}    }<\infty.
\]
Then the operator $VJ_{\Omega, n}$   is a trace  class  operator.
\end{corollary}

  Theorem~\ref{main1}  follows  from Lemma~\ref{lemma3.1} and  Corollary~\ref{corollary3.2}. $\,\,\,\,\Box$

\section{Proof of Theorem~\ref{main2}}

Our  study of the pure point spectrum is based on the  following  celebrated result of Simon and Wolff \cite{SW}:
\begin{theorem}
Let $A$ be a self-adjoint operator in a Hilbert  space ${\frak H}$   such that $\phi$ is a  cyclic  vector for $A$. Let $\mu_\phi$
be the spectral measure of $A$  corresponding to the vector $\phi$.  Let $P$ be the orthogonal projection onto the  space of scalar multiples of $\phi$.
 The   spectrum of $A+tP$  is  pure point on a Borel set  $\Omega\subset {\Bbb R}$
for almost every $t\in {\Bbb R}$,
if and  only if
\begin{equation}\label{SW}
\int_{{\Bbb R}}\frac{d\mu_\phi(t)}{(t-\lambda)^{2}}<\infty,\qquad  \text{for almost every} \quad \lambda\in \Omega.
\end{equation}
\end{theorem}

In applications to  discrete  Schr\"odinger operators,  it  is convenient to   interpret   \eqref{SW} as
the condition  that $(A-\lambda)^{-1}\phi$  is also an   element  of the Hilbert space  ${\frak H}$.
Indeed, let us define $U: {\frak H}\to L^2({\Bbb R},\mu_\phi)$ as a  bounded operator  mapping
an element of the form $f(A)\phi$  to  the function $f(t)$ (Here, one needs to consider only such functions  that $\phi\in {\mathcal D}(f(A))$).  This operator $U$  is a
  unitary operator from ${\frak H}$  onto $ L^2({\Bbb R},\mu_\phi)$.  Moreover, $U$  diagonalizes  the operator$A$ in the sense that   $UAU^{-1}$
is the operator of multiplication by the independent variable $t$.  Note now that $U\phi$  is the  function  that is  identically equal to 1.  Consequently, using  the fact that   the operator  $(A-\lambda)^{-1}$  is well defined for any $\lambda$  that is not an eigenvalue of $A$,  we obtain:
\[
\frac1{t-\lambda}\in L^2({\Bbb R},\mu_\phi) \iff \lambda\notin \sigma_p(A) \quad \text{and}\quad  \phi\in {\mathcal D} \bigl((A-\lambda)^{-1}\bigr).
\]

Thus, the
Simon-Wolff theorem can be   reformulated as follows.

\begin{theorem}\label{SW3}
Let $A$ be a self- adjoint operator in a Hilbert  space ${\frak H}$   such that $\phi$ is a  cyclic  vector for $A$.  Let $P$ be the orthogonal projection onto the  space of scalar multiples of $\phi$.
 The   spectrum of $A+tP$  is  pure point on a Borel set $\Omega$
for almost every $t\in {\Bbb R}$,
if and  only if  for almost every $\lambda\in \Omega \setminus \sigma_p(A)$,
\begin{equation}\label{SW2}
\phi\in{\mathcal  D}\bigl((A-\lambda)^{-1}\bigr).
\end{equation}
\end{theorem}

One  is tempted   to say
that   this theorem could be  applied  to  Schr\"odinger operators on the lattice ${\Bbb Z}^d$,  because one can decompose  $H=-\Delta+V$ into the orthogonal sum of
operators  whose   spectra   have   multiplicities  equal to  1. However, the   arguments that allow us to use  Theorem~\ref{SW3} are  more complicated.
\begin{proposition}\label{pr2.1}
Let ${\frak H}_0$ be  an invariant  subspace  for the  operator $H=-\Delta+V$  that  is  orthogonal to  vectors $\chi_j$  for all $j\in $ supp$V$.
Then the spectrum of the restriction of $H$ to the subspace  ${\frak H}_0$ is contained  in $[0,4d]$.
\end{proposition}
{\it Proof.} The  statement is  obvious and  its proof is left as an exercise. $\,\,\,\Box$

\begin{proposition}\label{pr2.2} Let $j\in {\Bbb Z}^d$ be a fixed point of the lattice.
Assume that  \[\chi_j\in {\mathcal D}\bigl((H-\lambda)^{-1}\bigr)\] for almost every $\lambda\in (a,b)\setminus \sigma_p(H)$. Let ${\frak H}_0(t)$ be an   invariant  subspace  for the  operator $H+t\chi_j(\cdot, \chi_j)$
on which   the restriction of  this operator  has  coninuous spectrum  that is contained in $[a,b]$. Then $\chi_j$ is orthogonal to  ${\frak H}_0(t)$  for almost every $t\in {\Bbb R}$.
\end{proposition}

{\it Proof.} Let ${\frak H}_1$ be the span of  vectors $\{\chi_j, H\chi_j, H^2\chi_j,\dots\}$. The  subspace ${\frak H}_1$   is invariant  for the  operator  $H+t\chi_j(\cdot, \chi_j)$ whose   restriction
to ${\frak H}_1$    is an operator having pure point spectrum in $[a,b]$  for almost every $t$. Consequently, ${\frak H}_0(t)$  is orthogonal to ${\frak H}_1$. $\,\,\,\Box$

\bigskip

\begin{corollary} Let $\{\xi_n\}_{n\in {\Bbb Z}^d}$ be independent  random variables  uniformly distributed on some interval $[-\alpha,0]$ where $\alpha>0$.
Let $\Omega\subset {\Bbb Z}^d$  be a  fixed subset of the lattice, such that
\[
V_\xi(n)=\begin{cases}
\xi_n,\qquad \text{if} \qquad n\in \Omega,\\
0,\qquad \text{if} \qquad n\notin \Omega.
\end{cases}
\]
Let $H_\xi=-\Delta+V_\xi$. Let also   $a<b<0$. Assume that,  for  all $\xi$  and all $j\in \Omega$, \[\chi_j\in {\mathcal D}\bigl((H_\xi-\lambda)^{-1}\bigr)\] for  almost every $\lambda\in (a,b)\setminus \sigma_p(H_\xi)$.
Then the operator  $H_\xi$ almost surely  has  pure point spectrum in $(a,b)$.
\end{corollary}

{\it Proof.} Let ${\frak H}_c(\xi)$  be the invariant  subspace on which the restriction of $H_\xi$  has  continuous spectrum  contained in $[a,b]$.
By Proposition~\ref{pr2.2},  $\chi_j$ is orthogonal to ${\frak H}_c(\xi)$  for almost  every $\xi_j\in[-\alpha,0]$ provided the  values of all other  random variables are fixed.
By Fubini's theorem,  $\chi_j$ is orthogonal to ${\frak H}_c(\xi)$  almost surely. By Proposition~\ref{pr2.1}, the   spectrum of the restriction of $H_\xi$  to ${\frak H}_c(\xi)$   does not intersect the interval $[a,b]$, because $b<0$.
The obtained  contradiction implies that a nontrivial subspace ${\frak H}_c(\xi)$  does not exist.
$\,\,\,\,\,\Box$

\bigskip

Let $H=-\Delta+V$ be a Schr\"odinger operator on the  lattice ${\Bbb Z}^d$.
Our  goal is to find   conditions guaranteeing   that  the sequence
\[
\psi(n)=\bigl[(H-\lambda)^{-1}\chi_j\bigr](n)
\]
is  square summable  for  any fixed $j\in {\Bbb Z}^d$. This litteraly means that   $\chi_j\in {\mathcal D}\bigl((H-\lambda)^{-1}\bigr)$.
We remind  the reader that, for $\lambda\notin [0,4d]$,  the  sequence $\psi$   is a  solution of the   equation
\begin{equation}\label{eqpsi}
\psi=\psi_0-(H_0-\lambda)^{-1}V\psi,
\end{equation}
where $\psi_0$  is the sequence
\[
\psi_0(n)=\bigl[(H_0-\lambda)^{-1}\chi_j\bigr](n)
\]
One the other hand,  once  we  know  the values $\psi(n)$  for $n\in $ supp$(V)$, we immediately obtain $\psi$  for all $n$, from \eqref{eqpsi}.

\begin{equation}\label{eqpsi2}
\psi(n)=\psi_0(n)-\Bigl((H_0-\lambda)^{-1}\chi_n,\chi_n\Bigr)V(n)\psi(n)-\sum_{l\neq n}\Bigl( (H_0-\lambda)^{-1}\chi_l,\chi_n\Bigr)V(l)\psi(l),
\end{equation}
First, observe that, for any $\varepsilon>0$, there are constants $\gamma_\varepsilon>0$ and $C_\varepsilon>0$  depending  on $\varepsilon$  such that \begin{equation}\label{expokernel}
\Bigl|\bigl( (H_0-\lambda)^{-1}\chi_l,\chi_n\bigr)\Bigr|\leq C_\varepsilon e^{-\gamma_\varepsilon|n-l|},\qquad  \forall n,l\in {\Bbb Z}^d,\qquad \forall\lambda \in {\Bbb R}\setminus [-\varepsilon,4d+\varepsilon].\end{equation}
\begin{proposition} Let $V$  be a bounded potential  and let $\varepsilon>0.$ Then
 for  almost every $\lambda\notin [-\varepsilon,4d+\varepsilon]$, there  is a  number $k(\lambda)$  such that
\begin{equation}\label{1+V}
\Bigl|1+\bigl((H_0-\lambda)^{-1}\chi_{n},\chi_{n}\bigr)V(n)\Bigr|\geq \frac 1 {|n|^{-d-\varepsilon}}
\end{equation}
for $|n|>k(\lambda)$.
\end{proposition}
{\it Proof.} Observe first, that if $\bigl|\bigl((H_0-\lambda)^{-1}\chi_{n},\chi_{n}\bigr)V(n)\bigr|<1/2$  and    $|n|>2$, then  \eqref{1+V} holds.
Threfore, we only need to consider the case $\bigl|\bigl((H_0-\lambda)^{-1}\chi_{n},\chi_{n}\bigr)V(n)\bigr|\geq 1/2$.
The latter implies that   we only need to consider   the points $\lambda\in [-2\|V\|_\infty,-\varepsilon]\cup [4d+\varepsilon, 4d+2\|V\|_\infty]$ and points $n\in {\Bbb Z}^d$ for which  $|V(n)|\geq \varepsilon/2$. On that set of $\lambda$-s,
the absolute value of the  derivative of the monotone  function
\[
f(\lambda)=1+\bigl((H_0-\lambda)^{-1}\chi_{n},\chi_{n}\bigr)V(n)
\]
is  bounded  below by $\varepsilon/\|V\|^2_\infty$. Therefore, the Lebesgue measure  of  the set
\[
S_\varepsilon(n):=\Bigl\{\lambda\in {\Bbb R}\setminus[-\varepsilon,4d+\varepsilon]:\qquad
\Bigl|1+\bigl((H_0-\lambda)^{-1}\chi_{n},\chi_{n}\bigr)V(n)\Bigr|<\frac 1 {|n|^{d+\varepsilon}}\Bigr\}
\]
is bounded  by
\[
|S_\varepsilon(n)|\leq \varepsilon^{-1}\|V\|^2_\infty|n|^{-d-\varepsilon}.
\]
Consequently,
\[
\sum_{n\in {\Bbb Z}^d}|S_\varepsilon(n)|<\infty.
\]  Using the Borel-Cantelli lemma,
we  conclude that, for  almost every  $\lambda\in [-\|V\|_\infty,-\varepsilon]\cup [4d+\varepsilon, 4d+\|V\|_\infty]$, there  is a  number $k(\lambda)$  such that
\[
\Bigl|1+\bigl((H_0-\lambda)^{-1}\chi_{n},\chi_{n}\bigr)V(n)\Bigr|\geq \frac 1 {|n|^{-d-\varepsilon}}
\]
for $|n|>k(\lambda)$. $\,\,\,\,\Box$

\bigskip

The equation \eqref{eqpsi2} can be  written in the form
\begin{equation}\label{eqpsi3}
\psi(n)=\alpha(n)\psi_0(n) -\alpha(n)\sum_{l\neq n}\Bigl( (H_0-\lambda)^{-1}\chi_l,\chi_n\Bigr)V(l)\psi(l),
\end{equation}
where the  sequence
\[
\alpha(n)=\bigl(1+\bigl((H_0-\lambda)^{-1}\chi_{n},\chi_{n}\bigr)V(n)\bigr)^{-1}
\]
obeys  the  bound
\begin{equation}\label{alphaocenka}|\alpha(n)|\leq |n|^{d+\varepsilon}\quad  \text{  for }\quad |n|>k(\lambda).
 \end{equation}
Consequently, \eqref{eqpsi3}  can be written in the form
\begin{equation}\label{eqpsi4}
\psi=\tilde\psi_0 -T\psi,
\end{equation} where $\tilde \psi(n)=\alpha(n)\psi(n)$  and $T$  is  the operator  defined  by
\begin{equation}\label{opT}
[Tu](n)=\alpha(n)\sum_{l\neq n}\Bigl( (H_0-\lambda)^{-1}\chi_l,\chi_n\Bigr)V(l)u(l),
\end{equation}
Combining \eqref{expokernel}  with \eqref{alphaocenka}, we conclude that  $T$ is  a compact operator on $\ell^2({\rm supp}(V))$ provided
the quantity \begin{equation}\label{definitiond}d(n)=\text{dist}\,\bigl(n, \,\, \text{supp}(V)\setminus \{n\}\bigr)\end{equation}
  obeys the sparseness condition
\begin{equation}\label{sparsen}
\lim_{ |n|\to\infty, \, n\in\text{supp}(V) }\frac{d(n)}{|n|^\delta}=\infty,\qquad \text{for some}\,\,\delta>0.
\end{equation}
\begin{proposition} \label{compact}Let $V$ be a bounded potential    such  that
\eqref{sparsen}  holds. Let $\chi_{{\rm supp}(V)}$  be the characteristic function of the support of  $V$.
 Let $T$  be defined by \eqref{opT}. Then the operator $\chi_{{\rm supp}(V)}T$  is compact.
\end{proposition}

{\it Proof.} For any $R>0$, let $\chi_{B_R}$ be the characteristic  function of the  set $B_R=\{n\in {\Bbb Z}^d:\,\, |n|\leq R\}$.
Consider the  decomposition
\[
T=\chi_{B_R}T+(1-\chi_{B_R})T.
\]
Since  $\chi_{B_R}T$ is a finite rank operator,   it is  sufficient to prove that
\[
\|(1-\chi_{B_R})\chi_{{\rm supp}(V)}T\|\to 0,\qquad {\rm as}\quad R\to \infty.
\]
The latter property follows  from the estimate
\begin{equation}\label{Shur}
\|(1-\chi_{B_R})\chi_{{\rm supp}(V)}T\|\leq   C_\varepsilon  \tilde C\sup_{n\in {\rm supp}(V):\,|n|>R} \bigl( |n|^{d+\varepsilon}(\sum_{l:\,|l- n|\geq d(n)}    e^{-\gamma_\varepsilon |n-l|} )^{1/2}   \bigr),
\end{equation}
which follows  from the Schur estimate  for the norm of an operator $K$ whose   matrix   consists of the  elements $k(n,m)=\rho_1(n,m)\rho_2(n,m)$.
This estimate says  that
\[
\|K\|\leq \sup_n \bigl( \sum_m|\rho_1(n,m)|^2\bigr)^{1/2}\sup_m \bigl( \sum_n|\rho_2(n,m)|^2\bigr)^{1/2}.
\]
The constant $\tilde C$ in \eqref{Shur}  is  equal to the quantity  $(\sum_{n\in {\Bbb Z}^d}    e^{-\gamma_\varepsilon |n|} )^{1/2}$.
$\,\,\,\,\Box$

\bigskip

Proposition~\ref{compact} implies that either \eqref{eqpsi4}  is  uniquely solvable, or the  equation
\begin{equation}\label{nontrivial}
\psi=-T\psi
\end{equation}
 has a non-trivial solution. On the other hand, since \eqref{nontrivial} is equivalent
to the equation
\[
\psi=-(H_0-\lambda)^{-1}V\psi,
\]
which  can be written in the  form $H\psi=\lambda \psi$,  the  equation \eqref{nontrivial} has a non-zero solution  if and  only if $\lambda$ is an eigenvalue of the operator $H$.
Consequently, \eqref{eqpsi4}  has a  unique  square summable solution $\psi$  for almost every $\lambda\notin [0,4d]$.
Thus, we obtain the following result.
\begin{theorem} Let $V$ be   bounded. Assume that  the quantity \eqref{definitiond} obeys
\[
\lim_{ |n|\to\infty, \, n\in \text{supp}(V) }\frac{d(n)}{|n|^\delta}=\infty
\]
for some $\delta>0$.
Then the sequence
\[
\psi(n)=\bigl[(H-\lambda)^{-1}\chi_j\bigr](n)
\]
is  square summable  for almost  every $\lambda\notin [0,4d]$.
\end{theorem}

\bigskip

As a  consequence, we obtain the following result.

\begin{theorem}\label{notmain} Let the  conditions of Theorem~\ref{main2} be fulfilled. Then
 the operator  $H_\xi$     almost surely  has  pure point spectrum on  ${\Bbb R}\setminus[0,4d]$.
\end{theorem}

Let us now explain why   the spectrum of the operator $H_\xi$ in  Theorem~\ref{main2} fills the interval $[\lambda_0,0]$.
\begin{theorem}\label{3.4}
Let $V$ be a bounded  potential for which the quantity \eqref{definitiond} obeys the condition
\[d(n)\to\infty,\qquad \text{as}\quad |n|\to\infty.
\]
Assume that there  is a  sequence of ponts $n_j\in{\Bbb Z}^d$
such that
\[
V(n_j)\to \beta<0,\qquad \text{as}\quad j\to\infty.
\]
Let $\lambda<0$ satisfy the equation
\[
1+\beta\bigl((H_0-\lambda)^{-1}\chi_0,\chi_0\bigr)=0.
\]
Then $\lambda\in \sigma(-\Delta+V)$.
\end{theorem}

{\it Proof.}
To prove Theorem~\ref{3.4}, we use the  three  statements below.

\begin{proposition}
Let $H=-\Delta+V$ and $H_n=-\Delta+V_n$ be Schrodinger operators on the lattice ${\Bbb Z}^d$  with bounded  potentials $V$ and $V_n$ correspondingly.
Assume that the sequence $V_n$ converges  to $V$ pointwise  so that
\[
\sup_{n\in {\Bbb N}} \|V_n\|_\infty<\infty.
\]
Then the sequence of measures
\[
\mu_n(\cdot)=\bigl((E_{H_n}(\cdot)\chi_0,\chi_0)\bigr)
\]
converges weakly to  the measure
\[
\mu(\cdot)=\bigl((E_{H}(\cdot)\chi_0,\chi_0)\bigr),\qquad \text{as}\quad |n|\to\infty.
\]
\end{proposition}

{\it Proof.} Observe that
\[
\int_{\Bbb R}\frac{d\mu(t)}{t-z}=\bigl((H-z)^{-1}\chi_0,\chi_0\bigr),\qquad z\in {\Bbb C}_+=\{z\in {\Bbb C}:\quad{\rm Im}\,z>0 \},
\]
and a similar representation can be written for  the measure $\mu_n$.
Since the span of  functions of the  form
\[
f(t)={\rm Im}\,\frac1{t-z},\qquad z\in {\Bbb C}_+,
\]
is  dense in the space of continuous decaying at infinity   functions $C_0({\Bbb R})$, it is  sufficient to show that
\[
\bigl((H_n-z)^{-1}\chi_0,\chi_0\Bigr)\to \bigl((H-z)^{-1}\chi_0,\chi_0\bigr),\qquad \text{as}\quad n\to\infty,
\]
uniformly on compact subsets  of ${\Bbb C}_+$. The latter  follows  from the Hilbert identity:
\[
\bigl((H_n-z)^{-1}\chi_0,\chi_0\bigr)-\bigl((H-z)^{-1}\chi_0,\chi_0\bigr)=\bigl((H_n-z)^{-1}(V-V_n)(H-z)^{-1}\chi_0,\chi_0\bigr).
\] $\,\,\,\,\,\Box$

\bigskip

\begin{corollary}\label{col3.2} Let $H=-\Delta+V$ be the  Schr\"odinger operator with a
bounded  potential  $V$  for which the quantity  \eqref{definitiond} obeys the condition
\[
d(n)\to\infty,\qquad \text{as}\quad |n|\to\infty.
\]
Let $n_j\in {\Bbb Z}^d$ be a sequence of distinct points  such that
\[
V(n_j)\to \beta<0,\qquad \text{as}\quad j\to\infty.
\]
Then the sequence of measures
\[
\mu_j(\cdot)=\bigl(E_{H}(\cdot)\chi_{n_j},\chi_{n_j}\bigr)
\]
converges   weakly to  the measure
\[
\mu(\cdot)=\bigl(E_{-\Delta+\beta P_0}(\cdot)\chi_0,\chi_0\bigr),
\]
where $P_0=(\cdot,\chi_0)\chi_0$  is the operator of multiplication by the function $\chi_0$.
\end{corollary}

{\it Proof.} It is  sufficient to note that
\[
\bigl(E_{H}(\cdot)\chi_{n},\chi_{n}\bigr)=\bigl(E_{H_n}(\cdot)\chi_0,\chi_0\bigr)
\]
where $H_n=-\Delta+V_n$  with $V_n(x)=V(x-n)$.  $\,\,\,\,\Box$

\bigskip

\begin{corollary}\label{col3.3}
Let  the conditions of Corollary~\ref{col3.2}  be fulfilled. Let $\lambda<0$ satisfy the equation
\[
1+\beta\bigl((H_0-\lambda)^{-1}\chi_0,\chi_0\bigr)=0.
\]
Then $\lambda\in \sigma(H)$.
\end{corollary}

{\it Proof.} Indeed, since the spectrum of $H$ contains  the support  of  the measure $\mu_n$  for all $n\in {\Bbb Z}^d$,
we conclude that
\[
\sigma(H)\supset {\rm supp} \bigl(E_{-\Delta+\beta P_0}(\cdot)\chi_0,\chi_0\bigr) \ni {\lambda}.
\]
$\,\,\,\,\,\Box$

\bigskip

Now Theorem~\ref{3.4}   follows  from  Corollary~\ref{col3.3}. $\,\,\,\,\Box$

\bigskip

\bigskip

Finally, we use the   following obvious  assertion.

\begin{proposition}\label{pro3.6}
Let $\Omega$ be an unbounded subset of the  lattice ${\Bbb Z}^d$. Let $\{\xi_n\}_{n\in \Omega}$ be  independent random variables uniformly distributed on $[-a,0]$ with some $a>0$.
Then, for any $\beta\in [-a,0]$ and  almost surely, there is a sequence of distinct points $n_j\in  \Omega$  such that
\[
\xi_{n_j}\to \beta,\qquad \text{as}\quad j\to\infty.
\]
\end{proposition}

{\it Proof.} Let   $B_R=\{n\in {\Bbb Z}^d:\,\,|n|\leq R\}$, where $R>0$.  It is enough to observe that for every $\varepsilon>0$ and every $R>0$, the probability of the event
\[
\{\xi_n\notin (\beta-\varepsilon,\beta+\varepsilon),\qquad  \forall n\in \Omega\setminus B_R\}
\]  is zero.  Consequently, there is at least one $n_1\in \Omega\setminus B_R$  for which $\xi_{n_1}\in (\beta-\varepsilon,\beta+\varepsilon)$. Continuing inductively, we construct a  sequence
$n_j$ such that  $\xi_{n_j}\in (\beta-2^{-j+1}\varepsilon,\beta+2^{-j+1}\varepsilon)$. $\,\,\,\,\Box$

\bigskip

Theorem~\ref{notmain}, Theorem~\ref{3.4} and Proposition~\ref{pro3.6} imply   Theorem~\ref{main2}.
\\
\\
\noindent {\bf \large Acknowledgments}:
The work of  S. Molchanov was supported by
the Russian Science Foundation,  project ${\rm N}^o$~20-11-20119.
The work of B. Vainberg was supported by the Simons Foundation grant 527180.

\end{document}